\documentclass[seceq]{ptptex}
\usepackage{graphicx}
\usepackage{amsfonts}
\usepackage{amssymb}
\usepackage{latexsym}
\usepackage{color}
\usepackage{cancel}
\input{colordvi.tex}

 \newcommand{\CL}{{\cal L}}

\newcommand{\bear}{\begin{array}} \newcommand{\eear}{\end{array}}
\newcommand{\bea}{\begin{eqnarray}} \newcommand{\eea}{\end{eqnarray}}
\newcommand{\beq}{\begin{equation}} \newcommand{\eeq}{\end{equation}}
\newcommand{\bef}{\begin{figure}} \newcommand{\eef}{\end{figure}}
\newcommand{\bec}{\begin{center}} \newcommand{\eec}{\end{center}}
\newcommand{\non}{\nonumber} 
\newcommand{\lmk}{\left(} \newcommand{\rmk}{\right)}
\newcommand{\lkk}{\left[} \newcommand{\rkk}{\right]}

\newcommand{\del}{\partial} 

\newcommand{\bib}{\bibitem}

\def\AAA#1#2#3{Astron. Astrophys. {\bf #1} (20#3), #2}

\def\APJJ#1#2#3{Astrophys. J. {\bf #1} (20#3), #2}

\def\APJSS#1#2#3{Astrophys. J. Suppl. {\bf #1} (20#3), #2}

\def\NPB#1#2#3{Nucl. Phys. B {\bf #1} (19#3), #2}

\def\PLB#1#2#3{Phys. Lett. B {\bf #1} (19#3), #2}

\def\PLBold#1#2#3{Phys. Lett. {\bf#1B} (19#3), #2}

\def\PRD#1#2#3{Phys. Rev. D {\bf #1} (19#3), #2}
\def\PRDD#1#2#3{Phys. Rev. D {\bf #1} (20#3), #2}

\def\PRL#1#2#3{Phys. Rev. Lett. {\bf#1} (19#3), #2}

\def\PTP#1#2#3{Prog. Theor. Phys. {\bf #1} (19#3), #2}

\title{Electroweak Baryogenesis with Embedded Domain Walls}

\author{Robert H. Brandenberger,$^{1,2}$ Wessyl Kelly,$^{2}$ and
Masahide Yamaguchi$^{3}$}
\inst{$^{1}$Physics Department, McGill University, Montr\'eal,
Qu\'ebec, H3A 2T8, Canada \\
$^{2}$Physics Department, Brown University, Providence, RI
02912, USA \\
$^{3}$Department of Physics and Mathematics, Aoyama Gakuin
University, Sagamihara 229-8558, Japan}

\recdate{November 24, 2006}

\abst{ We consider electroweak baryogenesis mediated by embedded domain
walls. Embedded domain walls originating from a symmetry breaking phase
transition are stabilized by thermal plasma effects, so that the
electroweak symmetry is unbroken in their cores. For this reason, the
cosmological evolution of such domain walls can generate a sufficiently
large baryon asymmetry, irrespective of the order of the electroweak
phase transition. For embedded domain walls, the condition that the
energy of the universe not be dominated by the energy of the domain
walls is relaxed significantly, and it is shown to be compatible with
our scenario of electroweak baryogenesis.}

\begin{document}

\maketitle

\section{Introduction}

\label{sec:intro}

The origin of the baryon asymmetry in the universe is one of the
greatest mysteries of modern cosmology and particle physics.  The
magnitude of this asymmetry has been determined by two methods.  One
method is Big Bang nucleosynthesis, which yields $\eta \equiv
n_{B}/n_{\gamma} = (3.4 - 6.9) \times 10^{-10}$ \cite{BBN}, where
$n_{B}$ is the net baryon number density and $n_{\gamma}$ is the photon
number density. The other method involves the study of the angular power
spectrum of cosmic microwave background anisotropies (which is effective
because changing the baryon number density changes the height of the
acoustic peaks in the spectrum), which yields $\eta = (6.14 \pm 0.25)
\times 10^{-10}$ \cite{WMAP}. These two estimates are consistent within
a 95\% confidence level.

Electroweak baryogenesis is one of the most promising scenarios to
explain the net baryon asymmetry, because it requires very little
physics beyond the Standard Model (SM) of particle physics. In
particular, as first emphasized in Ref. \citen{KRS}, by embedding the SM
in standard Big Bang cosmology, it is possible to satisfy Sakharov's
conditions for successful baryogenesis: (1) the existence of baryon
number violating processes; (2) that these processes violate {\it C} and
{\it CP} symmetries; and (3) the existence of deviations from thermal
equilibrium \cite{Sakharov}. In particular, the baryon number violating
processes in the SM are generated through the chiral anomaly.

Unfortunately, the minimal SM cannot produce a sufficiently large baryon
asymmetry, because the degree of {\it CP} violation is too small and, in
addition, within this model it is difficult to realize large deviations
from thermal equilibrium (see, e.g., Ref. \citen{BGrevs} for recent
reviews of electroweak baryogenesis). Thus, to realize electroweak
baryogenesis an extension of the SM must be considered. Sufficient {\it
CP} violation can be obtained, for example, by considering two-Higgs
models which explicitly contain {\it CP} violating terms in the Higgs
mass matrix in addition to the small amount of {\it CP} violation in the
fermion mass matrix present in the SM.

The nucleation and expansion of critical bubbles --- which are produced
if the phase transition is strongly first order --- are processes out of
thermal equilibrium. In the SM, the phase transition is not strongly
first order \cite{gleiser,MY}, and it has been pointed out that in fact
the electroweak phase transition is absent in the SM with experimentally
allowed Higgs masses \cite{KLRS}. However, a sufficiently strong phase
transition can be realized by considering supersymmetric extensions of
the SM. But in the minimal extension, severe conditions are required: In
particular, the Higgs bosons must have masses just above the present
experimental bounds and the stop quark must be lighter than the top
quark \cite{LR}.

Another scenario satisfying the out-of-equilibrium condition is to
consider topological defects in the cores of which the electroweak
symmetry is unbroken \cite{BD}. Specifically, in Ref. \citen{BD} the
role of electroweak strings \cite{EWstring} (which are non-topological
solitons arising in the SM and its extensions) was explored. There, it
was shown that the contraction of the loops of such strings generates a
departure from thermal equilibrium, just as does the expansion of a
bubble wall. In theories which admit defects, defects are formed for any
order of the phase transition. Thus, the out-of-equilibrium state is
realized even if the electroweak phase transition is of second
order. However, electroweak strings are unstable for realistic values of
the Weinberg angle \cite{JPV}.

As an extension of the above idea, one can consider cosmic strings
originating from a symmetry breaking that takes place before the
electroweak phase transition \cite{BDT}. If the electroweak symmetry is
restored and sphaleron processes are unsuppressed in the defect cores,
then electroweak baryogenesis is possible. Initially \cite{BDT}, this
baryogenesis scenario was investigated while taking into account only
local processes. In such studies, it was shown that baryogenesis is most
efficient when the new symmetry breaking scale is just above the
electroweak scale. In this case, the density of strings present at the
electroweak symmetry breaking scale is largest. Later, the analysis was
improved by including decay and nonlocal effects, which makes the
defect-mediated mechanism more efficient \cite{BDPT}. Nonetheless, the
amplitude of the baryon asymmetry which could be produced this way still
lies below the observed value. The main reason for the ineffectiveness
of this mechanism is that cosmic strings are one-dimensional objects,
and therefore, even if the strings are moving at relativistic speeds,
they sweep out too small a volume of space. An additional necessary
constraint \cite{Moore} is that the defects must be thick enough that
the electroweak sphalerons fit into their cores.

In contrast to cosmic strings, a scaling network of domain walls (see,
e.g., Refs. \citen{VilShell,HK,RHBrev} for reviews of topological defects
in cosmology) moving at relativistic speeds will sweep out a large
fraction of the universe. Thus, a sufficiently large baryon asymmetry
can be easily generated by making use of domain walls, provided that the
microscopic Lagrangian contains a sufficiently large {\it CP} violation
\cite{BDPT}.  However, domain walls produced at energy scales comparable
to or higher than the electroweak scale quickly dominate the energy
density of the universe. This causes severe cosmological problems. Thus,
in order to make domain wall-mediated electroweak baryogenesis viable,
we must introduce {\it ad hoc} processes that remove domain walls at
some time after the electroweak symmetry breaking but before the time of
nucleosynthesis. For example, a slight tilt is often introduced into the
scalar field potential, which breaks the degeneracy of the vacua such
that domain walls decay.

In this paper, we consider another method for wall-mediated electroweak
baryogenesis which avoids the domain wall problem. Instead of stable
domain walls, we make use of {\it embedded domain walls}. These have the
nice feature that their energy density decreases with the expansion of
the universe. The embedded defects \cite{ED} of a field theory are
solutions of the classical field equations whose energy distribution is
similar to that of a topological defect. However, in contrast to
topological defects, they are not topologically stable and thus decay in
the vacuum. However, if some of the fields are fixed, then the field
configuration becomes a topological defect in the constrained space.

It was recently argued \cite{NB} that many types of embedded defects can
be stabilized by interactions with a thermal plasma.  To illustrate this
effect, let us consider a multi-component real scalar field with a
``Mexican hat'' type potential. We consider the epoch after spontaneous
symmetry breaking when the scalar field is no longer in thermal
equilibrium.  In the case that all scalar field components except one
are charged, the charged fields acquire thermal masses from gauge
interactions with the photon bath. (Photons are still in thermal
equilibrium.) If the thermal mass is sufficiently large, then the
symmetry is broken only in the direction of the neutral scalar field. In
this way, embedded domain walls are produced and stabilized. If the
square of the induced plasma mass exceeds the absolute value of the
negative square mass of the scalar field at the top of its potential,
the symmetry is restored in the center of the defect. As the temperature
decreases further, the thermal masses also decrease, so that a core
phase transition occurs \cite{CPT}.  This means that the symmetric state
of the charged scalar fields in the core evolves into a slightly
asymmetric state, without destroying the overall defect. Thus, the
domain wall configurations survive even for small thermal
masses. However, the properties and cosmological evolution of these
domain walls differ from those of the conventional domain walls, as
shown below. The conditions necessary to avoid the cosmological domain
wall problem are thereby relaxed significantly.

In the next section, we investigate the properties and cosmological
evolution of embedded domain walls. Next, electroweak baryogenesis
mediated by these defects is investigated. In the final section, we give
discussion and conclusions.

\section{Embedded domain wall formation and dynamics}

Let us consider a multi-component real scalar field $\phi_{i}$ ($i = 1,
\cdots, N$) charged under a gauge group G, which breaks into a gauge
group H that includes the gauge group of the SM as a subgroup. The
Lagrangian density $\CL$ is given by
\beq
  \CL = \frac12 (D_{\mu} \phi_{i}) (D^{\mu} \phi_{i}) - V(\phi_{i})
        - \frac14 F^{a}_{\mu\nu} F^{a\,\mu\nu}, 
\eeq
where 
\beq
  F^{a}_{\mu\nu} = \del_{\mu} A^{a}_{\nu} - \del_{\nu} A^a_{\mu}
                   + g f_{abc} A^{b}_{\mu} A^{c}_{\nu} 
\eeq
is the field strength, $A^{a}_\mu$ is a gauge field, $g$ is a gauge
coupling constant, $D_{\mu} \phi_{i}$ is a covariant derivative defined
as $D_{\mu} \phi_{i} = (\del_{\mu} - i g A^{a}_{\mu} T^{a}) \phi_{i}$,
and $T^{a}$ is a generator of the gauge group G. Furthermore, we assume
a zero-temperature effective potential for these scalar fields of the
Coleman-Weinberg type \cite{CW},
\beq
  V_{\rm CW}(\phi_i) = \frac{\lambda}{4}\,\phi^4
     \lmk \ln \left| \frac{\phi}{v} \right| - \frac14 \rmk
     + \frac{\lambda}{16}v^4,
  \label{eq:CW}
\eeq
where $\phi = \sqrt{\Sigma \phi_i^2}$ and $\lambda$ is a coupling
constant. This potential can be obtained by considering a massless
scalar theory with one-loop corrections. Note that $V_{\rm CW}(\phi_i)$
also possesses global $O(N)$ symmetry. That is, in the case that gauge
interactions are negligible, the theory has approximate $O(N)$ symmetry
and can accommodate embedded domain wall solutions. Below, we consider
gauge interactions that break this global symmetry.

At high temperature, the components $\phi_i$ acquire finite temperature
corrections and the symmetry is restored. At the critical temperature,
the symmetry is broken. After a series of phase transitions, only
$U(1)_{\rm em}$ remains unbroken. Here, we further assume that one
component, $\phi_n$, is neutral under the would be $U(1)_{\rm em}$; the
other components, $\phi_{c1}, \phi_{c2}, \cdots$, are charged. In such a
situation, only the charged scalar fields acquire thermal masses,
$V_{T}(\phi_c)$,\footnote{We thank Guy D. Moore for informing us that
$V_{T}(\phi_c)$ changes for large $\phi_c$ due to the decoupling of
heavy particles.} originating from gauge interactions with an
electromagnetic coupling constant, $e$:\footnote{In fact, for charged
fields, gauge interactions also generate a zero-temperature effective
potential, which breaks the symmetry possessed by the potential
(\ref{eq:CW}). However, the global minimum of such a potential can
remain zero, even though the field values may be slightly changed at
that point. Furthermore, when we consider the configuration of domain
walls, this term changes the coefficient of the potential (\ref{eq:CW})
at $\phi_n=0$ only slightly, and the resulting effect is smaller than
these thermal effects.}
\begin{displaymath}
  V_{T}(\phi_c) = 
    \left\{
      \begin{array}{rl}
        \displaystyle{\frac12 e^2 T^2 \phi_c^2} & 
            \quad \mbox{for $\phi_c \lesssim T/e$}, \\ [0.4cm]
        \displaystyle{\frac{\pi^2 T^4}{45}} & 
            \quad \mbox{for $\phi_c \gtrsim T/e$},
      \end{array} 
    \right.
\end{displaymath}
where $\phi_c = \sqrt{\Sigma \phi_{cj}^2}$. Then, the total effective
potential becomes $V_{\rm eff}(\phi_i) = V_{\rm CW}(\phi_i) +
V_{T}(\phi_c).$ Note that this asymmetry between the neutral scalar
field and the charged scalar fields originates from the gauge
interactions, which break the global $O(N)$ symmetry possessed by the
potential term. Thus, this theory admits an embedded domain wall
solution based on the approximate $O(N)$ symmetry in which the charged
scalar fields vanish and the neutral one takes the form of the usual
single scalar field domain wall solution. In this domain wall solution,
the vacuum expectation value takes the form
$(\phi_{c1},\phi_{c2},\cdots,\phi_n)=(0,0,\cdots,\pm v)$, except inside
the cores of the walls, even after the core phase transition
\cite{NB}. The electroweak symmetry is kept unbroken inside the defect
cores by introducing interactions with the electroweak Higgs field $H$
of the form $h H^2 (\phi^4 - v^4) / v^2$, where $h$ is a coupling
constant.\footnote{The electroweak Higgs field may also be embedded in
the $\phi$ fields, for example, in a more complicated model, in which
the electroweak symmetry is automatically unbroken inside the cores,
just as in the standard GUT model.}
 
We now consider this model in a cosmological context. At very high
temperatures, the symmetry is unbroken, and the time averages of all
scalar fields vanish. Then, as the temperature decreases, the symmetry
is broken in the direction of the neutral component $\phi_n$ at some
critical temperature $T_{c}$, and (embedded) domain walls are formed.
As long as the photon field remains in thermal equilibrium, i.e. before
the time of the last scattering, these embedded walls persist, due to
the finite plasma masses for the charged scalar fields.

In order to investigate the dynamics of the charged components $\phi_c$,
we temporarily set $\phi_n = 0$. Then, the local maximum of the
potential in the direction of the charged components is given by
\beq
  \phi_{c,{\rm max}} \sim \frac{e T}{\sqrt{\lambda}} \, ,
\eeq
up to logarithmic corrections. Then, the global minimum is given by
\beq
  \phi_{c,{\rm min}} \sim v,
\eeq
with the potential energy
\beq
  V_{\rm eff}(\phi_{c,{\rm min}}) \sim \frac{\pi^2 T^4}{45}.
\eeq
Here we have assumed 
\beq
  T \ll e\,v \quad {\rm and} \quad e^4 \ll \lambda.
  \label{eq:pot}
\eeq

The width of the domain wall is determined by the balance between the
potential energy and the surface tension, and is given by
\beq
  \delta_{b} \sim \frac{1}{\sqrt{\lambda}\,v},
\eeq
which yields the following energy per unit area $\sigma_{b}$ of the
domain wall:
\beq
  \sigma_{b} \sim \sqrt{\lambda}\,v^3.
\eeq

After some relaxation period, the wall dynamics obey the ``scaling
solution''. In the scaling regime, the typical curvature radius of the
string network grows with the Hubble radius,\footnote{Such scaling
properties have been confirmed for local strings \cite{ls}, global
strings \cite{gs}, and global monopoles \cite{gm}. In our case, the
tension of the domain walls changes with the cosmic time. It has not yet
been confirmed that the scaling property holds for such walls. For this
reason, we need to investigate this topic further. It has already been
shown, however, that a cosmic string network with a time-dependent
tension {\it does} obey the scaling law.\cite{time}} and a few domain
walls of Hubble size will exist per Hubble volume \cite{DW}. Then, the
energy density of the domain wall network before the core phase
transition can be roughly estimated to be
\beq
  \rho_{b,{\rm DW}} \sim c_b\,\frac{\sigma_b t^2}{t^3}
       \sim c_b\,\frac{\sigma_b}{t},
\eeq
where $c_b$ is a constant of order unity. Then note that the total
energy density of the universe is given by
\beq
  \rho_{\rm total} \sim d\,\frac{M_G^2}{t^2},
\eeq
where $d$ is a number of order unity and $M_G \simeq 2.4 \times
10^{18}$~GeV is the reduced Planck scale. Thus, the time $t_{\rm eq}$ at
which domain walls begin to dominate the energy density of the universe
is given by
\beq
t_{\rm eq} \sim M_G^2 / \sigma_b \sim M_G^2 / (\sqrt{\lambda}\,v^3),
\eeq
up to numerical coefficients, which corresponds to
the temperature 
\beq
T_{\rm eq} \sim \sqrt{\sqrt{\lambda}\,v^3 / M_G} \, . 
\eeq
We have assumed that the universe is in the radiation-dominated epoch.

Though the effective mass squared at the origin is always positive, the
potential barrier decreases as the universe expands, and thus it can be
overcome through thermal fluctuations below some critical temperature
$T_{\rm tran}$. Then, a core phase transition takes place \cite{CPT}.
Note that, even after this core phase transition, the domain wall
configurations persist, due to the small thermal masses, even though
$\phi_n \ne 0$. However, the properties of the walls are changed
significantly. For example, the width of the wall after the core phase
transition is given by
\beq
  \delta_{a} \sim \frac{v}{T^2},
\eeq
and the energy per unit area $\sigma_{a}$ of the wall becomes
\beq
  \sigma_{a} \sim v T^2 \propto T^2.
\eeq
Note that $\sigma_{a}$ decreases in proportion to the temperature
squared. Thus, the energy density of such a domain wall network is
estimated to be
\[
   \rho_{a,{\rm DW}} \sim c_a \frac{\sigma_a}{t}
                    \sim c_a v \frac{T^2}{t}
                    \propto
    \left\{
      \begin{array}{rl}
        \displaystyle{a^{-4}} & 
            \quad \mbox{in the radiation-dominated universe}, \\ [0.3cm]
        \displaystyle{a^{-\frac72}} & 
            \quad \mbox{in the matter-dominated universe},
      \end{array} 
    \right.
\]
where $c_a$ is a constant of order unity. Therefore, embedded domain
walls after core phase transitions never dominate the energy density of
the universe as long as
\beq
  v \ll M_G.
  \label{eq:OD}
\eeq
Thus, $v$ can be larger than the electroweak scale without encountering
an overabundance problem. In previous papers \cite{NB,CBD,BDC}, it was
assumed that embedded defects stabilized by thermal plasmas decay after
recombination. However, the scalar fields have effective masses as long
as the root mean squared of the photon fields is nonzero, whether or not
these fields actually interact. Thus, though a more detailed analysis is
needed, we conjecture that the effective masses will not disappear and
the embedded defects will remain stable even after recombination. If the
embedded domain walls disappear after recombination, there will be no
domain wall problem, but even if they do not, the severity of this
problem is significantly lessened, as we have shown.

Note that if the embedded domain walls do not disappear after
recombination, the $U(1)_{\rm em}$ symmetry is broken inside the walls,
and this induces photon masses. Then, a photon wave which hits the
embedded wall can no longer propagate as a free wave. The mass profile
in the equation of motion of photons will induce nontrivial transmission
and reflection, so that CMB photons may be affected slightly. However,
today there will only be one wall per Hubble radius. We will not detect
the mass in our local experiments. None the less, effects from the
reflected photons might be detected. We will analyze this topic in the
future.

In order to find the critical temperature $T_{\rm tran}$ at which the
core phase transition takes place, we estimate the transition rate. This
transition occurs through bubble nucleation due to thermal
fluctuations. The typical radius $r_b$ of a bubble (strictly speaking,
the thickness of the bubble wall) is given by the curvature of the
potential at the top of the potential barrier, which is estimated as
\beq
  r_b \sim \frac{1}{ \sqrt{ \left| \frac{\del^2 V_{\rm eff}}{\del\phi_c^2}
                     (\phi_{c, {\rm max}}) \right| } }
      \sim \frac{1}{eT}.
\eeq
In the high temperature approximation, the transition rate $\Gamma$ is
found by considering the three-dimensional Euclidean action $S_3$
\cite{Linde},
\bea
  \Gamma \propto \exp \lmk - S_E \rmk,
\eea
where the Euclidean action $S_E$ is approximated as
\bea
  S_E &=& \frac{S_3}{T} \non \\
      &=& \frac{1}{T} \int d^3 x 
            \lkk \frac12 \lmk \nabla \phi \rmk^2 
            + V_{\rm eff}(\phi_c) - V_{\rm eff}(0) \rkk.
\eea
Minimizing $S_3$ gives the transition rate.

However, in our case, the transition takes place only inside the cores
of domain walls, and the typical radius $r_b$ is larger than $\delta_b$.
Thus, we should consider a two-dimensional action $S_2$ inside the cores
rather than the three-dimensional action $S_3$. Then, the Euclidean
action is given by
\beq
  S_{E} = \frac{1}{T} S_3
        = \frac{\delta_{b}}{T} S_{2},
\eeq
with
\beq
  S_{2} =  \int d^2 x 
           \lkk \frac12 \lmk \nabla\phi_c \rmk^2 
                  + V_{\rm eff}(\phi_c) - V_{\rm eff}(0) \rkk.
\eeq

In order to minimize $S_2$, we consider a circularly symmetric Gaussian
profile given by
\beq
  \phi_c \equiv \phi_{0} \exp \lmk - \frac{r^2}{R^2}
\rmk,
\eeq
which, as we will see below, is justified. Here $r$ is a radial
coordinate, $R$ is a typical radius, and $\phi_0 > 0$ is assumed for
simplicity. Then, the two-dimensional action $S_2$ becomes
\beq
  S_2 \simeq 2\pi \lkk \frac{\phi_0^2}{4}
              + \frac{\lambda}{32} \phi_0^4 R^2 
                 \lmk \ln \frac{\phi_0}{v} - \frac12 \rmk
              + \frac18 e^2 T^2 \phi_0^2 R^2
             \rkk,
\eeq
where we have also assumed that $\phi_0 \ll T/e$, which we also show is
justified below. Varying $S_2$ with respect to $\phi_0$ and $R$ yields
\bea 
  \frac{\del S_2}{\del \phi_0} &=& 2 \pi 
          \lkk \frac{\phi_0}{2}
              + \frac{\lambda}{8} \phi_0^3 R^2 
                 \lmk \ln \frac{\phi_0}{v} - \frac14 \rmk
              + \frac14 e^2 T^2 \phi_0 R^2
          \rkk, \non \\
  \frac{\del S_2}{\del R} &=& 2 \pi 
          \lkk \frac{\lambda}{16} \phi_0^4 R
                \lmk \ln \frac{\phi_0}{v} - \frac12 \rmk
              + \frac14 e^2 T^2 \phi_0^2 R
          \rkk.
\eea
Then $S_2$ is minimized for $\phi_0 \sim eT / \sqrt{\lambda} \ll T/e$
and $R \sim 1/(eT)$ and has a value of $S_2 \sim e^2 T^2 / \lambda$,
which yields the following Euclidean action $S_E$:
\beq
  S_E \sim \frac{\delta_b}{T} S_2 
      \sim \frac{e^2 T}{\lambda^{\frac32} v}. 
\eeq
Thus, the transition temperature $T_{\rm tran}$ becomes
\beq
 T_{\rm tran} \sim \frac{\lambda^{\frac32} v}{e^2},
 \label{eq:ttrans}
\eeq 
with a logarithmic correction coming from the prefactor of the
transition rate. Because $r_b \sim R$, the thick wall (Gaussian) ansatz
is justified. Note, again, that after the core phase transition, the
electroweak symmetry is broken even inside the cores of domain walls, so
that electroweak baryogenesis stops.

For successful baryogenesis, $T_{\rm tran}$ should be lower than the
electroweak scale $T_{\rm EW} \sim 100$~GeV, to prevent the electroweak
symmetry from breaking even inside the cores at that temperature. This
is the temperature at which electroweak baryogenesis is most effective.
Furthermore, $T_{\rm tran}$ should be higher than $T_{\rm eq}$, the
temperature at which the walls with symmetric cores would start to
dominate the energy density; otherwise the walls would dominate the
universe, and their subsequent decay would produce too large an entropy,
significantly diluting the baryon asymmetry and distort the black-body
nature of the CMB \cite{BDC}. These conditions yield the following
constraints on the breaking scale $v$:
\beq
  v \ll \frac{\lambda^{\frac52}}{e^4} M_{G},
\quad
  v \ll \frac{e^2}{\lambda^{\frac32}} T_{\rm 
EW}.
  \label{eq:bs}
\eeq
Note that condition (\ref{eq:pot}) should be satisfied at least at the
temperature $T_{\rm tran}$. Together (\ref{eq:pot}) and
(\ref{eq:ttrans}) give the constraint
\beq
  e^4 \ll \lambda \ll e^2.
  \label{eq:coupling}
\eeq
Under this constraint, the two conditions in (\ref{eq:bs}) are easily
satisfied. Thus, we have a large parameter region in which the
conditions (\ref{eq:OD}), (\ref{eq:bs}), and (\ref{eq:coupling}) are
satisfied.

\section{Electroweak baryogenesis}

Assuming that the criteria derived at the end of the previous section
are satisfied, we now investigate electroweak baryogenesis. In our
scenario, the embedded domain walls play the role of nucleated bubble
walls in standard electroweak baryogenesis. Other ingredients, such as
{\it CP} asymmetry, are assumed to be the same as in the standard case.

First, we consider local baryogenesis, in which the baryon asymmetry is
produced only at the edges of domain walls, because they are the locus
where {\it CP} violation in the scalar field sector is
localized. Defining the core passage time $\tau \equiv \delta_b / v_D$,
where $v_D$ is the typical wall velocity, the net baryon asymmetry after
decay is given by (see e.g. \citen{BDPT})
\beq
  n_B^{l} = n_B^{l,0} (1 - e^{-\overline{\Gamma}_s
\tau}),
\eeq
where $n_B^{l,0}$ is the baryon (anti-baryon) number density produced at
either edge of the defect. Here, $\overline{\Gamma}_s$ is defined as
$\overline{\Gamma}_s \equiv 6 N_f \Gamma_s / T^3$, using the weak
sphaleron rate $\Gamma_s = \kappa (\alpha_w T)^4$, with $\kappa = 0.1
\sim 1$.\cite{sphaleron}\footnote{B\"odeker found a different dependence
on the weak coupling constant, $\Gamma_S \propto \alpha_w^5
\ln(1/\alpha_w) T^4$\cite{BG}. However, we assume the above form for
simplicity, because the additional dependence can be absorbed into the
numerical uncertainty $\kappa$.} The quantity $N_f$ is the number of
families, and $\alpha_w \sim 1 / 29$ is the weak coupling
constant. Considering a chemical potential originating from the one-loop
effect \cite{one}, the baryon to entropy ratio is given by \cite{BDPT}
\beq
  \frac{n_B^l}{s} \sim 3.9 \kappa \alpha_w^4 g_{\ast}^{-1}
                     \lmk \frac{m}{T} \rmk \Delta\theta_{CP}
                  (1 - e^{-\overline{\Gamma}_s \tau}) \times
({\rm SF}),      
\eeq
where $m$ is the mass of the particles that make the dominant
contribution to the chiral anomaly, $\Delta\theta_{CP}$ is the magnitude
of the {\it CP} violation (assumed to be of order unity), and $({\rm
SF})$ is the geometrical suppression factor, which represents the ratio
of the volume in which baryogenesis actually occurs to the total
volume. The entropy density is given by
\beq
  s = \frac{2\pi^2}{45} g_{\ast} T^3, 
\eeq
where $g_{\ast}$ is the effective number of degrees of freedom.
 
Next, we consider non-local baryogenesis, particularly focusing on the
case in which baryon number violation is driven by chemical potentials
for left-handed leptons \cite{BDPT,JPT}. For other cases, the argument
is essentially the same. Defining the diffusion root $\lambda_D \equiv
v_D / D_L$, with the diffusion constant $D_L \sim 1/(8 \alpha_w^2 T)$ for
leptons, the net baryon asymmetry is given by
\beq
  n_B^{nl} = n_B^{nl,0} (1 - e^{-\delta_b \lambda_D}),
\eeq
where $n_B^{nl,0}$ is the baryon number density produced in front of the
trailing edge. Solving the diffusion equation for the flux of
left-handed leptons, we obtain the baryon to entropy ratio,
\beq 
  \frac{n_B^{nl}}{s} \sim 0.2 \kappa \alpha_w^2 g_{\ast}^{-1}
                     \Delta\theta_{CP} \frac{1}{v_D}
                     \lmk \frac{m_l}{T} \rmk^2 \lmk \frac{m_H}{T}
\rmk  
                     \lmk \frac{\xi_L}{D_L} \rmk \times ({\rm
SF})       
                     \sim 10^{-6} \kappa \Delta\theta_{CP} 
                          \frac{y_{\tau}^2}{v_D} \times
({\rm SF}),      
\label{final}
\eeq
where $m_l$ is the mass of the leptons, $m_H$ is the Higgs mass, $\xi_L$
is the persistence length, and $y_{\tau}$ is the Yukawa coupling
constant for tau leptons. In the second step, we have used the fact that
contributions coming from tau leptons are dominant. In the case that
domain walls evolve for a long time according to the scaling solution,
the equilibrium baryon number is reached as long as the sphaleron is
active inside the cores, which enhances the baryon number $n_B^{nl}$ by
a factor of $v_D^2 / (\overline{\Gamma}_s D_L)$ \cite{BDPT}.

In our case, the suppression factor $({\rm SF})$ is given by the wall
velocity $v_D$, because domain walls are two-dimensional
objects. Therefore, as long as wall motion is relativistic, there is no
significant suppression, and hence [as shown by (\ref{final})] a
sufficiently large baryon asymmetry can easily be produced in our
scenario.

\section{Discussion and summary}

In this paper, we have studied electroweak baryogenesis in the context
of embedded domain walls. The walls are stabilized by thermal plasma
effects, so that the electroweak symmetry is restored and sphaleron
processes are active inside their cores. Then, as we have shown, a
sufficiently large baryon asymmetry can easily be produced during the
cosmological evolution of the wall network. In contrast to the case of
stable cosmic strings, the volume factor that suppresses defect-mediated
baryogenesis from a baryogenesis process which is effectively uniform in
space is not significant, because domain walls are two-dimensional
objects, which, when moving relativistically, sweep out a fraction of
order unity of space. This makes it easy to produce a sufficient baryon
asymmetry.

In general, a network of stable domain walls will easily dominate the
energy density of the universe, leading to a severe cosmological
problem. But, in our case, the domain walls are stabilized by thermal
plasma and undergo a core phase transition at some critical
temperature. This transition reduces their energy per unit area as the
tension of the domain walls decreases in proportion to the cosmic
temperature squared, and this leads to a significant relaxation of the
condition needed for the domain walls not to dominate the energy of the
universe. The walls may even decay completely after recombination.
 
Note that our scenario requires the use of a Coleman-Weinberg potential,
as opposed to a potential in which the scalar field has a positive mass
squared at the origin. In spite of this caveat, the positive results of
this investigation warrant further investigation of the fate of
embedded defects stabilized by thermal plasmas.  This is a topic of
future work.

\section*{Acknowledgments}

We are grateful to Alessio Notari, Balaji Katlai, Guy D. Moore, and
Natalia Shuhmaher for useful discussions concerning the plasma
stabilization of embedded defects.  R. B.\ is supported in part (at
McGill) by an NSERC Discovery grant and (at Brown) by the
U.S. Department of Energy under Contract DE-FG02-91ER40688,
TASK~A. M. Y.\ is supported in part by a JSPS Grant-in-Aid for
Scientific Research (No. 18740157) and by a project of the Research
Institute of Aoyama Gakuin University.

\end{document}